\documentclass{article}
\usepackage{spconf,amsmath,graphicx,amssymb}
\usepackage{amsfonts}
\usepackage{booktabs}
\usepackage{multirow}
\usepackage{hyperref}
\usepackage{siunitx}
\usepackage{mathtools,mathrsfs}
\usepackage[subrefformat=parens]{subcaption}
\usepackage{comment}
\usepackage[linesnumbered,ruled,vlined]{algorithm2e}
\usepackage{color}

\def\equationautorefname~#1\null{(#1\null)}

\let\oldnl\nl
\newcommand{\nonl}{\renewcommand{\nl}{\let\nl\oldnl}}
\newlength\mylen

\SetCommentSty{mycommfont}

\newcommand{\tablescale}{0.85}

\usepackage{xspace}
\makeatletter
\DeclareRobustCommand\onedot{\futurelet\@let@token\@onedot}
\def\@onedot{\ifx\@let@token.\else.\null\fi\xspace}
\def\eg{\emph{e.g}\onedot} 
\def\ie{\emph{i.e}\onedot}

\makeatother

\newcommand{\vect}[1]{\mbox{\boldmath $#1$}}
\newcommand{\abs}[1]{\left\lvert#1\right\rvert}

\newcommand{\trans}[1]{#1^\mathsf{T}}

\DeclareMathOperator*{\argmax}{arg\,max}
\def\appendixautorefname~#1\null{~#1 \null}
\newcommand{\relmid}{\mathrel{}\middle|\mathrel{}}

\makeatletter
\newcommand{\figcaption}[1]{\def\@captype{figure}\caption{#1}}
\newcommand{\tblcaption}[1]{\def\@captype{table}\caption{#1}}
\makeatother

\title{End-to-End Speaker Diarization as Post-Processing}

\name{
Shota Horiguchi$^{1}$ \qquad
Paola Garc\'{i}a$^{2}$ \qquad
Yusuke Fujita$^{1}$ \qquad
Shinji Watanabe$^{2}$ \qquad
Kenji Nagamatsu$^{1}$
}
\address{
$^{1}$ \text{Hitachi, Ltd. Research \& Development Group, Japan} \\
$^{2}$ \text{Center for Language and Speech Processing,
            Johns Hopkins University, USA}
}

\begin{document}
\ninept

\abovedisplayskip=6pt
\belowdisplayskip=6pt
\setlength\floatsep{13pt}
\setlength\textfloatsep{13pt}
\setlength\intextsep{13pt}
\setlength\abovecaptionskip{5pt}
\setlength\belowcaptionskip{0pt}
\setlength\dbltextfloatsep{13pt}

\aboverulesep=0.25ex 
\belowrulesep=0.5ex 
\maketitle
\begin{abstract}
This paper investigates the utilization of an end-to-end diarization model as post-processing of conventional clustering-based diarization.
Clustering-based diarization methods partition frames into clusters of the number of speakers; thus, they typically cannot handle overlapping speech because each frame is assigned to one speaker.
On the other hand, some end-to-end diarization methods can handle overlapping speech by treating the problem as multi-label classification. Although some methods can treat a flexible number of speakers, they do not perform well when the number of speakers is large.
To compensate for each other's weakness, we propose to use a two-speaker end-to-end diarization method as post-processing of the results obtained by a clustering-based method.
We iteratively select two speakers from the results and update the results of the two speakers to improve the overlapped region.
Experimental results show that the proposed algorithm consistently improved the performance of the state-of-the-art methods across CALLHOME, AMI, and DIHARD II datasets.
\end{abstract}
\section{Introduction}
\label{sec:intro}
Speaker diarization, which is sometimes referred to as ``who spoke when'', has important roles in many speech-related applications.
It is sometimes used to enrich transcriptions by adding speaker attributes \cite{anguera2012speaker}, and at the other times, it is used to improve the performance of speech separation and recognition \cite{boeddeker2018front,medennikov2020stc}.

Speaker diarization methods can be classified roughly into two: clustering-based methods and end-to-end methods.
Typical clustering-based methods i) first classify frames into speech and non-speech, ii) then extract an embedding which describes speaker characteristics from each speech frame, and iii) finally apply clustering to the extracted embeddings.
Most methods employ hard clustering such as agglomerative hierarchical clustering (AHC) and k-means clustering; as a result, each frame belongs either to one of the speaker clusters or to the non-speech cluster.
The assumption that underlies these clustering-based methods is that each frame contains at most one speaker, \ie, they treat speaker diarization as a set partitioning problem.
Thus, they fundamentally cannot deal with overlapping speech.
Despite the assumption, they are still strong baselines over end-to-end methods on datasets of a large number of speakers, \eg, DIHARD II dataset \cite{ryant2019second}.
This is because they handle multiple speaker problems based on unsupervised clustering without using any speech mixtures as training data.
Thus, the methods do not suffer from overtraining due to the lack of the overlap speech especially for a large number of speakers.

On the other hand, some end-to-end methods called EEND treat speaker diarization as a multi-label classification problem.
They predict whether each speaker is active or not at each frame; thus, they can deal with speaker overlap.
Evaluation of the early models fixed the number of speakers to two \cite{fujita2019end1,fujita2019end2,fujita2020endtoend}.
Some extensions are proposed recently to handle unknown number of speaker cases, \eg, encoder-decoder-based attractor calculation \cite{horiguchi2020endtoend} and one-by-one prediction using speaker-conditioned model \cite{fujita2020neural}.
However, these methods still perform poorly when the number of speakers is large. 
One reason is the training datasets. Mixtures of a large number of speakers are often rare in various datasets; thus, end-to-end models cannot produce diarization results for large number of speakers because they are overtrained on mixtures of a few number of speakers.
Even if the issue on the number of mixtures is solved, the EEND depends on the permutation invariant training \cite{yu2017permutation} so that it is still hard to train the model on a large number of mixtures in terms of the calculation cost.
For these reasons, how to handle mixtures that contain overlapping speech of a large number of speakers is still an open problem for both clustering and end-to-end diarization methods.

In this paper, we propose to combine both clustering-based and end-to-end methods effectively to deal with overlapping speech regardless of the number of speakers.
We first obtain the initial diarization result using x-vector clustering, which does not produce overlapping results in most cases.
We then apply the following steps iteratively: i) frame selection to contain only two speakers and silence and ii) overlap estimation using a two-speaker EEND model.
The frame selection is also used to adapt the EEND model to a dataset which contains mixtures of more than two speakers.
We evaluate our method using various datasets including CALLHOME, AMI, and DIHARD II datasets.

\section{Related work}
\subsection{Clustering-based diarization}
While some methods provide supervised clustering of speaker embeddings \cite{zhang2019fully}, the most common approach is an x-vector clustering in an unsupervised manner (See the systems submitted to DIHARD II Challenge, \eg, \cite{landini2020but,lin2020dihard}).
Since naive x-vector clustering results in poor performance, various techniques to improve the performance have been proposed, \eg, probabilistic linear discriminate analysis (PLDA) rescoring \cite{sell2014speaker} and Variational Bayes (VB) hidden Markov model (HMM) resegmentation \cite{diez2018speaker}.
In terms of overlap processing, most methods first detect overlapped frames and then assign the second speaker for the detected frames based on heuristics \cite{landini2020but,landini2019but} or the results of VB resegmentation \cite{diez2019bayesian}.

Another direction is based on clustering of overlapped segments \cite{huang2020speaker}.
It first extracts overlapped segments using a region proposal network, and then applies clustering for embeddings extracted from each of them.
It fundamentally solved the issue of embedding extraction using a sliding window, but its accuracies are not comparable to end-to-end methods described in \autoref{sec:endtoend}.

\subsection{End-to-end diarization for overlapping speech}
\label{sec:endtoend}
One end-to-end approach is called EEND \cite{fujita2019end1,fujita2019end2,fujita2020endtoend}. They calculate multiple speaker activities, each corresponding to a single speaker. Recent models can output a flexible number of speakers' activities by using encoder-decoder-based attractor calculation modules (EDA) \cite{horiguchi2020endtoend} or speaker-conditional EEND (SC-EEND) \cite{fujita2020neural}.
Another approach is called RSAN, which are based on residual masks in the time-frequency domain to extract speakers one by one \cite{kinoshita2018listening,kinoshita2020tackling}.

While EEND and RSAN take only acoustic features as input, a variant of these methods also accepts a speaker embedding as input to determine the target-speaker and output his/her speech activities.
For example, target-speaker voice activity detection (TS-VAD) uses i-vectors to output the corresponding speakers' voice activities \cite{medennikov2020targetspeaker_short}, but the number of speakers is fixed by the model architecture.
Personal VAD \cite{ding2020personal} and VoiceFilter-Lite \cite{wang2020voicefilterlite}, which are based on d-vectors, have not such a limitation, but they assume that each speaker's d-vector is stored in the database in advance; thus they are not suited for speaker-independent diarization.

\section{Proposed method}
\label{sec:proposed}
\subsection{Overview}
Given acoustic features $\{\vect{x}_t\}_{t=1}^T$, where $t\in\{1,\dots,T\}\eqqcolon[T]$ denotes a frame index, diarization is a problem to predict a set of active frames $\mathcal{T}_k\subseteq[T]$ for each speaker $k\in\{1,\dots,K\}\eqqcolon[K]$.
$K$ is the estimated number of speakers.
For simplicity, we use $\mathcal{X}_\mathcal{T}\coloneqq\{\vect{x}_t\mid t\in\mathcal{T}\}$ to denote the features of selected frames $\mathcal{T}\subseteq[T]$.

Clustering-based methods assume that input recordings do not contain speaker overlap. It formulates diarization as a set partitioning problem, \ie, $\mathcal{T}_k$ for $k\in[K]$ are predicted to be disjoint, \ie, $\forall\left\{i,j\right\}\in\binom{[K]}{2},\mathcal{T}_i\cap\mathcal{T}_j=\emptyset$.
In EEND, on the other hand, diarization is formulated as a multi-label classification to handle overlapping speech; thus, they do not have to be disjoint.
The formulation of EEND is appropriate for real conversations in which speakers sometimes utter simultaneously. 
However, it makes the problem too difficult to be solved; when $K$ is large (\eg 10), it rarely happens that $K$ speakers speak together.
Therefore, we assume that at most $K'(< K)$ speakers speak simultaneously, and refine the clustering-based results using an end-to-end model that is trained to process at most $K'$ speakers.
In this study, we set $K'=2$.
The detailed algorithm is explained in \autoref{sec:algorithm}.

\subsection{Algorithm}\label{sec:algorithm}
\begin{figure*}[t]
    \centering
    \includegraphics[width=\linewidth]{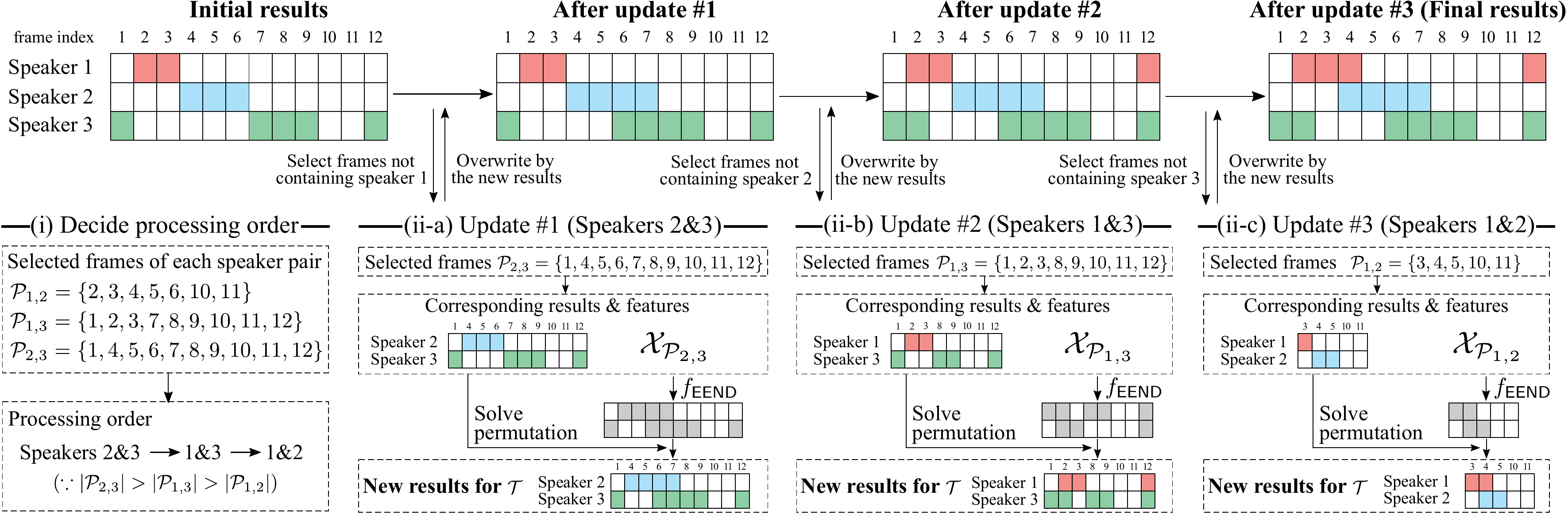}
    \caption{The flow of the proposed method when the number of speakers is three. Given initial diarization results (top left), our method (i) picks up a pair of speakers iteratively in decreasing order of the number of frames and (ii) refines the diarization results of the two speakers using an end-to-end speaker diarization model.}
    \label{fig:flow}
\end{figure*}

Given initial diarization results $\{\mathcal{T}_k\mid\emptyset\neq\mathcal{T}_k\subseteq[T]\}_{k=1}^{K}$, we iteratively select two speakers among $K$ and update the diarization results of the two speakers using an EEND model.
The EEND model $f_\mathsf{EEND}:\mathbb{R}^{D\times L}\rightarrow\left(0,1\right)^{2\times L}$ was trained to estimate posteriors probabilities of two speakers from an $L$-length sequence of $D$-dimensional acoustic features.
\autoref{fig:flow} show the flow of the proposed method when $K=3$.

\subsubsection{Processing order determination}
To apply the iterative refinement to each pair of speakers, the processing order influences the accuracy of final diarization results.
This is because we cannot select frames to include only two speakers based on estimated diarization results because they include diarization errors.
For example, if we select frames not containing Speaker 1 in \autoref{fig:flow}, the fourth frame contains Speaker 1 according to the final results.
If the ratio of such impurities among the selected frames is high, the refinement using EEND may not perform well.
We found that this problem is simply solved by processing the pairs of speakers in decreasing order of the number of selected frames (\autoref{fig:flow}(i)).
For each speaker pair $\{i,j\}\in\binom{[K]}{2}$, we first select a set of frames $\mathcal{P}_{i,j}$ not containing speakers other than $i$ and $j$ as follows:
\begin{align}
    \mathcal{P}_{i,j}&=[T]\setminus\bigcup_{k\in[K]\setminus\left\{i,j\right\}}\mathcal{T}_k.
    \label{eq:frame_selection}
\end{align}
We then apply the refinement described below for each speaker pair in descending order of $\abs{\mathcal{P}_{i,j}}$ as in \autoref{fig:flow} (ii-a)--(ii-c).

\subsubsection{Iterative update of diarization results}
To update the diarization results of speakers $i$ and $j$, we first reselect a set of frames $\mathcal{P}_{i,j}$ using \autoref{eq:frame_selection} .
This is because the diarization results $\{\mathcal{T}_k\}_{k=1}^{K}$ are updated at each refinement step so that we cannot reuse the one that is calculated to decide the processing order.
Then the corresponding features $\mathcal{X}_{\mathcal{P}_{i,j}}$ are input to the EEND model to obtain posteriors of two speakers $(a)$ and $(b)$ by
\begin{align}
    \left(\trans{\left[q_t^{(a)},q_t^{(b)}\right]}\relmid t\in\mathcal{P}_{i,j}\right)=f\left(\mathcal{X}_{\mathcal{P}_{i,j}}\right)\in\left(0,1\right)^{2\times\abs{\mathcal{P}_{i,j}}},
\end{align}
where $q_t^{(a)}$ and $q_t^{(b)}$ denote posteriors of the first and second speakers at frame index $t$, respectively, and $\trans{(\cdot)}$ denotes the matrix transpose.
We simply apply the threshold value of 0.5 to obtain the indexes of active frames of the two speakers as 
\begin{align}
    \mathcal{Q}^{(a)}=\left\{t\in\mathcal{P}_{i,j}\relmid q_t^{(a)}>0.5\right\},\;
    \mathcal{Q}^{(b)}=\left\{t\in\mathcal{P}_{i,j}\relmid q_t^{(b)}>0.5\right\}.
\end{align}
Note that we have speaker permutation ambiguity between $(a)$--$(b)$ and $i$--$j$, and we solve permutation to find the optimal correspondence between $(\mathcal{T}_i,\mathcal{T}_j)$ and $(\mathcal{Q}^{(a)},\mathcal{Q}^{(b)})$ as follows:
\begin{align}
    \left(\hat{\mathcal{T}}_i,\hat{\mathcal{T}}_j\right)&=\argmax_{(u,v)\in\{(a,b),(b,a))\}}s\left(\mathcal{Q}^{(u)},\mathcal{T}_i\right)+s\left(\mathcal{Q}^{(v)},\mathcal{T}_j\right),
\end{align}
where $s(\mathcal{U},\mathcal{V})$ is a function to calculate similarity between speech and non-speech activities described by two sets $\mathcal{U}$ and $\mathcal{V}$ defined as 
\begin{align}
    s(\mathcal{U},\mathcal{V})&\coloneqq\underbrace{\abs{\mathcal{U}\cap\mathcal{V}}}_{\text{speech similarity}}+\underbrace{\abs{\left([T]\setminus\mathcal{U}\right)\cap\left([T]\setminus\mathcal{V}\right)}}_{\text{non-speech similarity}}.
\end{align}
Finally, we update the diarization results of speakers $i$ and $j$.
To confirm that the new results $\hat{\mathcal{T}}_i$ and $\hat{\mathcal{T}}_j$ are calculated for speaker $i$ and $j$, we check whether they satisfy the following conditions:
\begin{align}
    \frac{\abs{\hat{\mathcal{T}}_i\cap\left(\mathcal{T}_i\cap\mathcal{P}_{i,j}\right)}}{\abs{\mathcal{T}_i\cap\mathcal{P}_{i,j}}}>\alpha,\quad
    \frac{\abs{\hat{\mathcal{T}}_j\cap\left(\mathcal{T}_j\cap\mathcal{P}_{i,j}\right)}}{\abs{\mathcal{T}_j\cap\mathcal{P}_{i,j}}}>\alpha,
    \label{eq:conditions}
\end{align}
where $\alpha$ is a lower limit of the ratio of the intersection between the new results $\hat{\mathcal{T}}_i$ (or $\hat{\mathcal{T}}_j$) and the previous results $\mathcal{T}_i\cap\mathcal{P}_{i,j}$ (or $\mathcal{T}_j\cap\mathcal{P}_{i,j}$).
In this study, we set $\alpha=0.5$.
Only if the conditions in \autoref{eq:conditions} are satisfied, we update the results of speakers $i$ and $j$.
When $K=2$, we simply update the results with the new ones as
\begin{align}
    \mathcal{T}_i\leftarrow\hat{\mathcal{T}}_i\cup\left([T]\setminus\mathcal{P}_{i,j}\right),\quad
    \mathcal{T}_j\leftarrow\hat{\mathcal{T}}_j\cup\left([T]\setminus\mathcal{P}_{i,j}\right).
    \label{eq:update_k2}
\end{align}
On the other hand, when $K\geq3$, such fully-update strategy causes a performance drop due to impurities in the selected frames.
Thus, we use the following instead of \autoref{eq:update_k2} to update only overlapped frames:
\begin{align}
    \mathcal{T}_i\leftarrow\mathcal{T}_i\cup\left(\hat{\mathcal{T}}_i\cap\hat{\mathcal{T}}_j\right),\quad
    \mathcal{T}_i\leftarrow\mathcal{T}_j\cup\left(\hat{\mathcal{T}}_j\cap\hat{\mathcal{T}}_j\right).
\end{align}

For the end-to-end model $f_\mathsf{EEND}$, we use the self-attentive EEND model with an encoder-decoder attractor calculation module (SA-EEND-EDA) \cite{horiguchi2020endtoend}.
It consists of a four-layer-stacked Transformer encoder to extract embeddings for each frame and the EDA module to calculate attractors from the extracted embeddings.
The EDA includes long short-term memories but we shuffled the order of embeddings just before they are fed into the EDA, which improves the diarization performance.
Thus, we can consider that all the components of $f_\mathsf{EEND}$ are independent of the order of embeddings and therefore the model can treat input features of selected frames even if they are not continuous in time.

\subsection{Training strategy of the SA-EEND-EDA model}
\label{sec:training_strategy}
In the original EEND and its derived methods \cite{fujita2019end2,horiguchi2020endtoend,fujita2020neural} used \textit{matched} dataset for model adaptation, \ie, only two-speaker subset of the original dataset (\eg. CALLHOME\footnote{\url{https://catalog.ldc.upenn.edu/LDC2001S97}}) was used to finetune the models in two-speaker evaluations.
This strategy cannot be used to finetune two-speaker models when the dataset does not contain two-speaker mixtures (\eg. AMI \cite{carletta2007unleashing}).
Even if two-speaker mixtures are included in the dataset, it does not make full use of the datasets, which may cause performance degradation.

To cope with this situation, we adopt the frame-selection technique used in \autoref{eq:frame_selection} for model adaptation.
If the input chunk contains more than two speakers, we first choose two dominant speakers and then eliminate frames in which the other speakers are active as in \autoref{eq:frame_selection}.
The model is trained only using the selected frames to output speech activities of the two speakers.
This makes it possible to finetune two-speaker models from any kind of multi-speaker datasets without mixture-wise selection.

\section{Experiment}
\label{sec:experiment}

\subsection{Settings}
\label{sec:dataset}
\autoref{tbl:dataset} shows the datasets used for our evaluation.
The model was pretrained using simulated two-speaker mixtures \textit{Sim2spk} for 100 epochs.
Each mixture was simulated from two single-speaker audios derived from Switchboard-2 (Phase I \& II \& III), Switchboard Cellular (Part 1 \& 2), or NIST Speaker Recognition Evaluation (2004 \& 2005 \& 2006 \& 2008).
Noise sources from MUSAN corpus \cite{snyder2015musan} and simulated room impulse responses \cite{ko2017study} are also used to simulate noisy and reverberant environments.
The detailed simulation protocol is in our previous paper \cite{fujita2019end2}.
For the pretraining, Adam optimizer with the learning rate scheduler proposed in \cite{vaswani2017attention} was used. The number of warm-up steps was set to 100,000 following \cite{horiguchi2020endtoend}.

After the pretraining, the model was adapted on CALLHOME, AMI \cite{carletta2007unleashing} and DIHARD II \cite{ryant2019second} datasets for another 100 epochs, respectively.
Adam optimizer was also used in the adaptations but its learning rate was fixed to $1\times10^{-5}$ following \cite{horiguchi2020endtoend}.

\begin{table}[t]
    \centering
    \setlength{\tabcolsep}{4.4pt}
    \caption{Dataset to train and test our diarization models.}
    \label{tbl:dataset}
    \vspace{8pt}
    \resizebox{\linewidth}{!}{
    \begin{tabular}{@{}llccc@{}}
        \toprule
        Dataset & &\#Speakers & \#Mixtures & Overlap ratio (\%)\\\midrule
        \textbf{Pretrain}&Sim2spk& 2 & 100,000&34.1\\\midrule
        \textbf{Adaptation}&CALLHOME-2spk & 2 & 155 & 14.0\\
        &CALLHOME&2--7&249&17.0\\
        &AMI train \cite{carletta2007unleashing}&3--5&118&19.4\\
        &DIHARD dev \cite{ryant2019second}&1--10&192&9.8\\\midrule
        \textbf{Test}&CALLHOME-2spk &2 & 148 & 13.1\\
        &CALLHOME &2--6 & 250 & 16.7\\
        &AMI eval \cite{carletta2007unleashing}&3--4&24&18.6\\
        &DIHARD eval \cite{ryant2019second} &1--9 &194&8.9\\
        \bottomrule
    \end{tabular}
    }
\end{table}

We used diarization error rate (DER) and Jaccard error rate (JER) for evaluation.
While some studies excluded overlapped regions from evaluation \cite{sell2014speaker,zhang2019fully}, this study scored overlapped region.
We also note that our evaluations are based on estimated speech activity detection (SAD), while some studies used oracle segments \cite{diez2019bayesian} or only reported confusion errors \cite{zhang2019fully}.

\subsection{Preliminary evaluation of the training using frame selection}
Before the evaluation of the proposed post-processing method, we first evaluated the training strategy explained in \autoref{sec:training_strategy} using two transformer-based two-speaker EEND models: SA-EEND \cite{fujita2019end2} and SA-EEND-EDA \cite{horiguchi2020endtoend}.
They were trained on CALLHOME-2spk in the original papers, but we utilized mixtures that contain more than two speakers in the CALLHOME dataset.
\autoref{tbl:results_callhome_twospk} shows DERs on the CALLHOME-2spk test set.
Using the full CALLHOME improved DER of SA-EEND from \SI{9.54}{\percent} to \SI{9.00}{\percent} and that of SA-EEND-EDA from \SI{8.07}{\percent} to \SI{7.84}{\percent}.
According to these results, we show the effectiveness of our training strategy described in \autoref{sec:training_strategy} based on the frame selection with \autoref{eq:frame_selection}.
\begin{table}[t]
    \centering
    \caption{DERs (\%) on CALLHOME-2spk. Collar tolerance of \SI{0.25}{\second} is allowed.}
    \scalebox{\tablescale}{
    \begin{tabular}{@{}lcc@{}}
        \toprule
        Model&Adaptation&DER\\\midrule
        SA-EEND \cite{fujita2020endtoend} &CALLHOME-2spk& 9.54\\
        SA-EEND &CALLHOME + frame selection& 9.00\\\midrule
        SA-EEND-EDA \cite{horiguchi2020endtoend} &CALLHOME-2spk& 8.07\\
        SA-EEND-EDA &CALLHOME + frame selection& \textbf{7.84}\\
        \bottomrule
    \end{tabular}
    }
    \label{tbl:results_callhome_twospk}
\end{table}

\subsection{Results}
\subsubsection{CALLHOME}
We first evaluated the proposed method on CALLHOME dataset, which is composed of telephone conversations.
As a clustering-based baseline, x-vectors with AHC and PLDA\footnote{\url{https://github.com/kaldi-asr/kaldi/tree/master/egs/callhome_diarization/v2}} was used with TDNN-based speech activity detection\footnote{\url{https://github.com/kaldi-asr/kaldi/tree/master/egs/aspire/s5}}.
We also prepared the results for which VB-HMM resegmentation \cite{diez2018speaker} was applied.
All the components were implemented in Kaldi recipe.

\autoref{tbl:callhome_results} shows the evaluation results.
X-vector clustering without and with VB achieved \SI{19.43}{\percent} and \SI{17.61}{\percent} DERs, respectively, but they didn't outperform the \SI{15.29}{\percent} DER scored by SA-EEND-EDA trained to output diarization results on flexible number of speakers.
However, we can also observe that the clustering-based methods are better when the number of speakers is larger than four.
Applying the proposed post-processing for x-vector clustering baselines achieved \SI{16.79}{\percent} and \SI{14.06}{\percent} without and with VB, and the latter is \SI{1.23}{\percent} better than the SA-EEND-EDA model.
In terms of the number of speakers, the proposed method performed well on both large and small number of speakers.

\begin{table}[tb]
    \centering
    \setlength{\tabcolsep}{4.2pt}
    \caption{DERs (\%) on CALLHOME. All the results include overlapped regions and are NOT based on oracle SAD. Collar tolerance of \SI{0.25}{\second} is allowed.}
    \label{tbl:callhome_results}
    \scalebox{\tablescale}{
    \begin{tabular}{@{}lcccccc@{}}
        \toprule
        &\multicolumn{5}{c}{\#Speakers}\\\cmidrule(l){2-6}
        Method &  2 & 3 & 4 & 5 & 6 & All\\\midrule
        SA-EEND-EDA \cite{horiguchi2020endtoend}& \textbf{8.50} & 13.24 & 21.46 & 33.16 & 40.29 & 15.29\\\midrule
        x-vector AHC&15.45&18.01&22.68&31.40&34.27&19.43\\
        x-vector AHC + Proposed&13.85&14.72&18.61&\textbf{28.63}&29.02&16.79\\\midrule
        x-vector AHC + VB&12.62&16.82&21.27&31.14&31.80&17.61\\
        x-vector AHC + VB + Proposed&9.87&\textbf{13.11}&\textbf{16.52}&28.65&\textbf{27.83}&\textbf{14.06}\\
        \bottomrule
    \end{tabular}
    }
\end{table}

\subsubsection{AMI}
Second, we evaluated our method on AMI dataset, consisting of meeting recordings. While it includes various types of recordings, we used \textit{Headset mix} recordings for this experiment.
We chose the system developed during JSALT 2019 \cite{garcia2020speaker_short} as a baseline.
It is based on x-vector clustering followed by VB resegmentation and overlap detection and assignment for the second speaker candidate \cite{bullock2020overlap}.

\autoref{tbl:ami_results} shows DERs and JERs on AMI eval set.
The proposed method reduced DERs of \SI{3.07}{\percent}, \SI{3.14}{\percent}, and \SI{0.18}{\percent} of absolute improvement from the three baselines.
Surprisingly, our method improved DER and JER of the results in which the overlap detection \cite{bullock2020overlap} was already applied.

\begin{table}[t]
    \centering
    \setlength{\tabcolsep}{4.4pt}
    \caption{DERs and JERs (\%) on AMI eval. VB: Variational Bayes resegmentation, OVL: Overlap detection and speaker assignment \cite{bullock2020overlap}. All the results include overlapped regions and are NOT based on oracle SAD. No collar tolerance is allowed.}
    \label{tbl:ami_results}
    \scalebox{\tablescale}{
    \begin{tabular}{@{}lcc@{}}
        \toprule
        Method&DER&JER\\\midrule
        x-vector AHC \cite{garcia2020speaker_short}& 33.75&45.68\\
        x-vector AHC + Proposed&30.64&43.78\\\midrule
        x-vector AHC + VB \cite{garcia2020speaker_short}&32.80&43.72\\
        x-vector AHC + VB + Proposed&29.66&42.63\\\midrule
        x-vector AHC + VB + OVL \cite{garcia2020speaker_short}&28.15&41.00\\
        x-vector AHC + VB + OVL + Proposed&\textbf{27.97}&\textbf{40.57}\\
        \bottomrule
    \end{tabular}
    }
\end{table}

\subsubsection{DIHARD II}
Finally, we evaluated the proposed method on DIHARD II dataset, which includes recordings from 10 different domains.
We used the official baseline system \cite{ryant2019second} and the BUT system \cite{landini2020but,landini2019but}, which is the winning system of the second DIHARD Challenge, to obtain initial diarization results.
Both are based on the x-vector clustering, but the BUT system is more polished in that it extracts x-vectors in shorter intervals and uses VB resegmentation and overlap detection and assignment based on heuristics.

The results are shown in \autoref{tbl:dihard_results}.
The proposed method reduced DER and JER of the baseline system by \SI{2.96}{\percent} and \SI{2.91}{\percent}, respectively.
Our method also improved DER and JER of \SI{0.35}{\percent} and \SI{0.66}{\percent} from the BUT system without overlap assignment and \SI{0.38}{\percent} and \SI{0.64}{\percent} from that with overlap assignment, respectively.
These improvements are small, but it is far better than the heuristic-based overlap assignment in \cite{landini2020but}, which improved DER by \SI{0.15}{\percent} ($=27.26-27.11$) and JER by \SI{0.08}{\percent} ($=49.15-49.07$).

\begin{table}[t]
    \centering
    \caption{DERs and JERs (\%) on DIHARD II eval. All the results include overlapped regions and are NOT based on oracle SAD. No collar torelance is allowed.}
    \label{tbl:dihard_results}
    \scalebox{\tablescale}{
    \begin{tabular}{@{}lcc@{}}
        \toprule
        Method&DER&JER\\\midrule
        DIHARD II baseline \cite{sell2018diarization}& 40.86&66.60\\
        DIHARD II baseline + Proposed&37.90&63.79\\\midrule
        BUT system (w/o OVL) \cite{landini2020but,landini2019but}& 27.26&49.15\\
        BUT system (w/o OVL) + Proposed & 26.91&48.49\\\midrule
        BUT system (w/ OVL) \cite{landini2020but,landini2019but}& 27.11&49.07\\
        BUT system (w/ OVL) + Proposed&\textbf{26.88}&\textbf{48.43}\\
        \bottomrule
    \end{tabular}
    }
\end{table}

\section{Conclusion}
\label{sec:conclusion}
In this paper, we proposed a post-processing method for clustering-based diarization using an end-to-end diarization model. We iteratively selected two speakers, picked up frames that contain the two speakers, and process the frames by the end-to-end model to update diarization results.
Evaluations on CALLHOME, AMI, and DIHARD II datasets showed that our proposed method improves various types of clustering-based diarization results.

\section{Acknowledgment}
We would like to thank Federico Landini for providing the results of the winning system \cite{landini2020but,landini2019but} of the second DIHARD Challenge.

\bibliographystyle{IEEEbib-abbrev}
\bibliography{refs}

\end{document}